\documentstyle[aps,epsf,floats,amsmath]{revtex}

\begin{document}
\draft
\parskip 2mm
\twocolumn[\hsize\textwidth\columnwidth\hsize\csname@twocolumnfalse%
\endcsname
\title{First order phase transition with a logarithmic singularity \\
in a model with absorbing states}
\author {Haye Hinrichsen}
\address{
Theoretische Physik, Fachbereich 10,
Gerhard-Mercator-Universit\"at Duisburg, D-47048 Duisburg,
Germany}

\date{August 10, 2000}
\maketitle
\begin{abstract}
Recently, Lipowski {[cond-mat/0002378]} investigated a stochastic 
lattice model which exhibits a discontinuous transition from an
active phase into infinitely many absorbing states. 
Since the transition is accompanied by an apparent 
power-law singularity, it was conjectured that the
model may combine features of first- and second-order phase transitions. 
In the present work it is shown that this singularity emerges as an 
artifact of the definition of the model in terms of products. 
Instead of a power law, we find a logarithmic 
singularity at the transition. Moreover, we generalize 
the model in such a way that the second-order phase transition 
becomes accessible. As expected, this transition belongs to the 
universality class of directed percolation.
\end{abstract}

]
%

\section{Introduction}

In nonequilibrium statistical physics the study of 
phase transition continues to attract considerable 
attention~\cite{MarroDickman99}. In this context, continuous
phase transition into absorbing states have been of particular
interest~\cite{HinrichsenReview}. It is believed that 
absorbing-state transitions
can be categorized into universality classes, the most prominent ones
being directed percolation (DP)~\cite{DP} and the so-called
parity-conserving~(PC) universality class~\cite{PC}. On the other
hand, various nonequilibrium models are known to
exhibit a discontinuous phase transition~\cite{FirstOrder}. 
Especially in one spatial dimension first-order transitions require
a very robust mechanism in order to stabilize the ordered phases. 
As suggested in Ref.~\cite{First1D}, first-order transitions 
in one dimension should be impossible under certain generic conditions
if one of the ordered phases fluctuates.

Recently, Lipowski introduced a model with infinitely many absorbing
states which exhibits a first-order transition from a fluctuating
active state into an absorbing phase~\cite{OriginalPapers,Lipowski}. 
Remarkably, this transition takes place even in one spatial dimension.
Unlike previously investigated models, the dynamic rules 
involve {\em products} of real-valued local variables. Surprisingly, 
the first-order transition is accompanied by an apparent power-law 
singularity of the stationary particle density, suggesting that the 
model may combine features of continuous and
discontinuous phase transitions. 
This observation collides with the commonly accepted belief that there are
no power-law singularities in first-order phase transitions. The aim of
this work is to study the origin of this unusual type of singularity
in more detail.

\begin{figure}
\epsfxsize=75mm
\centerline{\epsffile{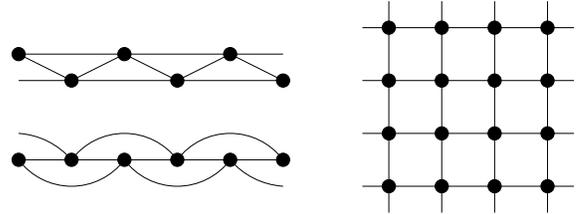}}
\vspace{2mm}
\caption{
\label{FIGMODEL}
Lattice geometries used in Ref.~[8].
Left: One-dimensional triangular ladder, interpreted as a linear
chain with next-to-nearest neighbor interactions. Right: Two-dimensional
square lattice.
}
\end{figure}

The model considered in Ref.~\cite{Lipowski} is defined as follows.
Each site $i$ of a given lattice is connected to $n$
neighboring sites $j \in <i>$. Each bond carries a real-valued 
variable $w_{ij}=w_{ji} \in (-1/2,+1/2)$. The set of all bond variables
specifies the state of the system. A site is considered to be active
if the product of all adjacent bond variables is smaller than
a certain control parameter $r$:
\begin{equation}
\label{ProductDefinition}
\prod_{j \in <i>} w_{i,j} < r \,.
\end{equation}
Initially all bond variables are uniformly distributed 
between $-1/2$ and $+1/2$. The model evolves by 
random sequential updates according to the following dynamic
rules. For each update attempt a site is randomly selected.
If it is active, all adjacent bond variables are replaced by
new random numbers distributed between $-1/2$ and $+1/2$.
Lipowski considered the case of $n=4$ neighbors using a one-dimensional
triangular ladder and a two-dimensional square lattice
(see Fig.~\ref{FIGMODEL}).

Due to the use of real-valued local variables,
the model has infinitely many absorbing
states. Moreover, the dynamic rules are invariant under certain
gauge transformations. For example, we may invert the sign of 
all bond variables along a closed contour without changing the 
pattern of activity in a given configuration. The implications of
this type of gauge invariance are not yet fully understood.

In order to understand the existence of a phase transition,
let us consider two extremal situations. On the one hand, for 
$r>2^{-n}$ it is obvious that all sites are active during the entire
temporal evolution. On the other hand, for $r<0$ there is a finite
probability to generate bond variables with  $|w_{i,j}|<2^{n-1}|r|$. 
This means that certain pairs of sites $i$ and $j$ remain inactive 
forever, irrespective of the values of the other bond variables. 
Thus, the process continuously "switches off" certain pairs of 
sites and therefore approaches an absorbing configuration within 
an exponentially short time. 

\begin{figure}
\epsfxsize=85mm
\centerline{\epsffile{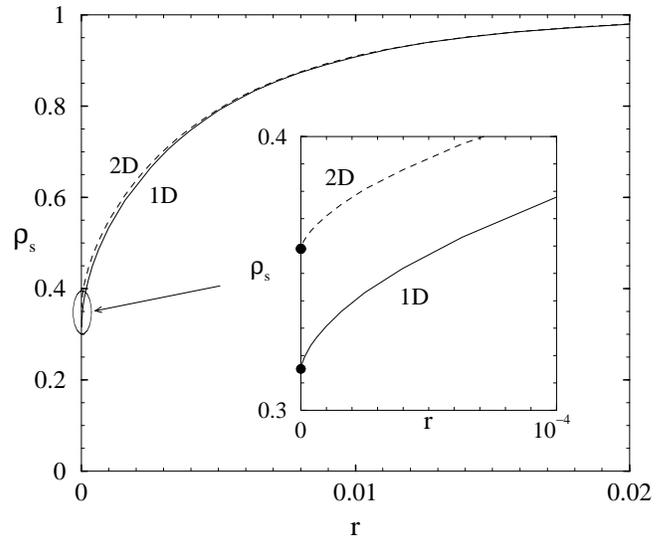}}
\vspace{2mm}
\caption{
\label{FIGDENS}
Stationary density $\rho_s$ as a function of the parameter~$r$ in
one (1D) and two (2D) spatial dimensions. The inset zooms
the region where the curves terminate. The terminal points at 
$\rho(0)=\rho_0$ are marked by the bold dots.
}
\end{figure}

Interestingly, for r=0 the model is still in the active phase
with a non-vanishing stationary density of active sites $\rho_0>0$ 
(see Fig.~\ref{FIGDENS}). Thus, the spreading process undergoes a 
{\em discontinuous} phase transition at $r=0$. 
Even more surprisingly, the stationary density $\rho_{s}(r)$
does not decrease linearly versus $\rho_0$ as $r \rightarrow 0$,
instead the slope of the curve seems to diverge. This observation
led Lipowski to the conjecture that the model may combine features
of discontinuous and continuous transitions, calling for a power-law
behavior of the form
\begin{equation}
\label{PowerLaw}
\begin{split}
\rho_{s}(r) &\;\simeq\; \rho_0 + a \, r^\beta \,
\qquad \text{for $r\geq 0$}\,, \\
\rho_{s}(r)  &\;=\;0  \hspace{20.6mm} \text{for $r<0$}\,.
\end{split}
\end{equation}
where $a$ is a certain factor and $\beta$ is the critical 
exponent associated with the order parameter $\rho$.
Performing numerical simulations Lipowski found the estimates
$\rho_0=0.314827, \beta=0.66(3)$ in one dimension
and $\rho_0=0.358, \beta=0.58(1)$ in two dimensions, respectively. 
Moreover, he observed that the dynamical critical exponent 
$z=\nu_\perp/\nu_\parallel \simeq 0.2$ is very small.

In the present paper we propose a different explanation of the
diverging slope in Fig.~\ref{FIGDENS} based on the following
arguments:
\begin{enumerate}
\item
The singularity of the slope in Fig.~\ref{FIGDENS} emerges as an
artifact of the definition of activity in terms of a product.
While the bond variables are uniformly distributed, the
probability distribution of the product diverges for 
$r \rightarrow 0$, leading to a singularity of the slope. 
An explicit formula is derived, showing that the slope 
diverges logarithmically as $r \rightarrow 0$. 
In particular, there there is no power law of the 
form~(\ref{PowerLaw}). 

\item Redefining the control parameter, the model 
displays a conventional first-order phase transition
without a diverging slope.
Moreover, there is no diverging length scale 
at the transition.

\item 
Since for $r<0$ the system is immediately driven towards one of the
absorbing states, the continuous 
transition may be thought of as being hidden in the inaccessible
region $r<0$. In order to support this point of view, 
we generalize the model in such a way that the continuous transition 
is shifted to the accessible region $r \geq 0$. As expected, 
the transition belongs to the universality class of directed percolation.  

\end{enumerate}
%
%

\section{Logarithmic singularity}

In this Section we demonstrate that the singularity of the slope
in Fig.~\ref{FIGDENS} is a consequence of the multiplicative
definition of activity in Eq.~(\ref{ProductDefinition}). To this
end we consider the {\it reactivation probability} 
$W(r)$ that a site remains active after an update. 
In a spreading process, the reactivation probability provides a
good measure of the effective spreading rate by which nearest
neighbors will be activated. In most spreading processes near
the transition, the reactivation probability varies
to lowest order linearly 
with  the control parameter of the model. Therefore, a power law
of the form $\rho_s \sim (W(r)-W(r_c))^\beta$ immediately implies the
same power law in terms of the control parameter 
$\rho_s \sim (r-r_c)^\beta$. In the present model, however, 
$W(r)$ is a {\em nonlinear} function at the transition. 
Therefore, it does matter whether $r$ or $W(r)$ is 
used as control parameter. 

Our explanation relies on the assumption that $W(r)$ is the "true"
control parameter of the model. In terms of $W(r)$ the model exhibits 
a regular first order phase transition without singularity, i.e.,
the density varies linearly with $W(r)-W(r_c)$.
The apparent singularity in terms of $r$ originates solely in the
non-analytic behavior of the function $W(r)$ in the limit
$r \rightarrow 0$. 

In order to compute $W(r)$ let us consider $n$ independent
random numbers $z_1,z_2,\ldots,z_n$ drawn from a flat distribution 
between $-1/2$ and $+1/2$. Clearly, the probability
$P^{(1)}(z)dz$ to find one of these random numbers between
$z$ and $z+dz$ is constant for $|z| \leq 1/2$. However, the product
$z=\prod_{i=1}^{n}z_i$ of the random numbers is not uniformly
distributed. The probability distribution $P^{(n)}(z)$ can be
computed iteratively by
\begin{equation}
\begin{split}
P^{(k+1)}(z)&=
\int_{-\frac12}^{+\frac12} dz'
\int_{-\frac{1}{2^k}}^{+\frac{1}{2^k}} dz''
\, P^{(k)}(z') \, \delta(z-z'z'') \\
&=
\int_{-\frac12}^{+\frac12} dz'\,
\frac{1}{z'} \, P^{(k)}(z/z')\, \Theta(z'-|z|2^k) \\
&=
2 \int_{-2^k|z|}^{+\frac12} dz'\,
\frac{1}{z'} \, P^{(k)}(|z|/z')\,
\end{split}
\end{equation}
with $P^{(1)}(z)=1$, leading to the exact result
\begin{equation}
P^{(n)}(z)=\frac{(-2)^{n-1}}{(n-1)!}\,(\log_e 2^n|z|)^{(n-1)}\,.
\end{equation}
Thus, the probability $W(r)$ to reactivate an updated site is
given by
\begin{equation}
\begin{split}
W(r) &= \int_{-\frac{1}{16}}^r dz \, P^{(4)}(z) \\& = 
\frac12+r\,
\Bigl(\,8 - 8 (\log_e 16 |r|) + \\ 
& \hspace{17mm} 4 (\log_e 16 |r|)^2  - \frac{4}{3}(\log_e 16 |r|)^3 \Bigr) \,.
\end{split}
\end{equation}
For small values of $r$, we therefore expect the
stationary density to be given by
\begin{equation}
\label{LogarithmicLaw}
\begin{split}
\rho_s(r) &\;\simeq\; \rho_0 \,+\, A \, \Bigl(W(r)-\frac12 \Bigr)
\qquad \text{for $r\geq 0$}\,, \\
\rho_s(r) &\;=\;0  \hspace{37mm} \text{for $r<0$}\,,
\end{split}
\end{equation}
where $A$ is a fit parameter. This formula replaces the
power law in Eq.~(\ref{PowerLaw}), implying that the slope
of $\rho_s(r)$ diverges {\em logarithmically} as $-(\log_e 16 |r|)^3$
for $r \rightarrow 0$ in any dimension. This explains why Lipowski's 
results in one and two dimensions are so similar.

In order to support the validity of Eq.~(\ref{LogarithmicLaw}), 
we performed Monte Carlo simulations (see Table~\ref{Tab}). 
Our estimate for the stationary density $\rho_0$ in one dimension
deviates slightly from the value quoted in~\cite{Lipowski}.
This deviation may be explained as follows. On the one hand,
the random number generator plays a crucial role. Since most 
algorithms generate internally an integer random number, 
the output is often quantized in steps of about $10^{-8}$, leading
to wrong results if $r$ is very small. 
On the other hand, the machine precision itself
limits the accuracy. This problem can be avoided by
storing $\log(|w_{ij}|)$ and $\text{sgn}(w_{ij})$ instead of
$w_{i,j}$ and turning the product in Eq.~(\ref{ProductDefinition})
into a sum of logarithms. Taking these technical subtleties into 
account, we obtain a different estimate.

\begin{figure}
\epsfxsize=85mm
\centerline{\epsffile{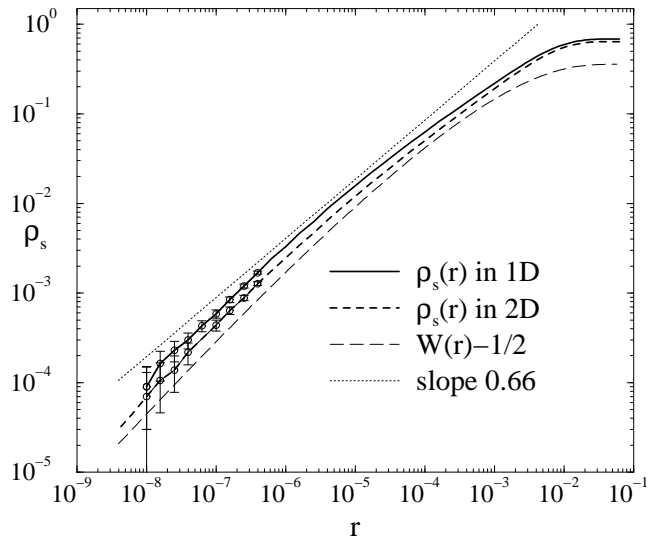}}
\vspace{2mm}
\caption{
\label{FIGLOGDENS}
Log-log plot of the stationary density $\rho_{s}(r)$ 
as a function of the parameter $r$, compared to the
(vertically shifted) function $W(r)-1/2$. The dotted straight line
visualizes the failure of the power-law conjecture 
proposed in~[8].
}
\end{figure}

\begin{figure}
\epsfxsize=85mm
\centerline{\epsffile{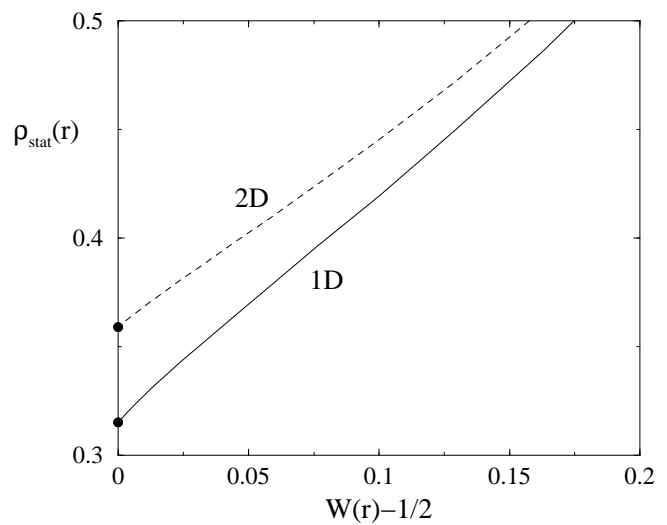}}
\vspace{2mm}
\caption{
\label{FIGTRUEDENS}
Stationary density $\rho_{s}$ as a function of $W(r)-1/2$.
As can be seen, there is no singularity near the transition.}
\end{figure}

As shown in Fig.~\ref{FIGLOGDENS}, the results for the stationary 
density in one and two dimensions are very similar. Moreover, for
small values of $r$ the curvature of the lines is in fair agreement 
with the theoretical prediction of Eq.~(\ref{LogarithmicLaw}).
In any case, the possibility of a power-law singularity can be ruled out.
As expected, there is no singularity if $\rho_{s}$ is plotted
against $W(r)-W(0)$ (see Fig.~\ref{FIGTRUEDENS}).

\begin{table}
\begin{tabular}{|c||c|c|}
 & $d=1$ & $d=2$ \\
\hline
\hline number of sites    & $10^6$ & $1000^2$ \\
\hline simulation time  & $10^4\ldots 10^6$ & $10^4\ldots 10^6$  \\
\hline  \ \ stationary density $\rho_0$  \ \ & 0.31512(3) \ \ & 0.35905(2) \ \ \\
\hline fit parameter $A$ & 1.4(1) & 1.1(1)
\end{tabular}
\vspace{2mm} 
\caption{\label{Tab} Numerical estimates.}
\end{table}

%
%

\section{Continuous transition in a generalized model}

As outlined in the Introduction, the first-order phase transition
at $r=0$ is induced by frozen pairs of inactive sites. Since for
any negative value of $r$ there is a finite probability 
to generate such pairs during the temporal evolution, 
the system is driven into one of the absorbing states
within an exponentially short time. Roughly speaking, the spreading
process is switched off as soon as $r<0$. 
Thus, the continuous transition, which is expected to exist in
ordinary spreading processes, may be thought of as being 
hidden in the inaccessible region $r<0$. 

\begin{figure}
\epsfxsize=85mm
\centerline{\epsffile{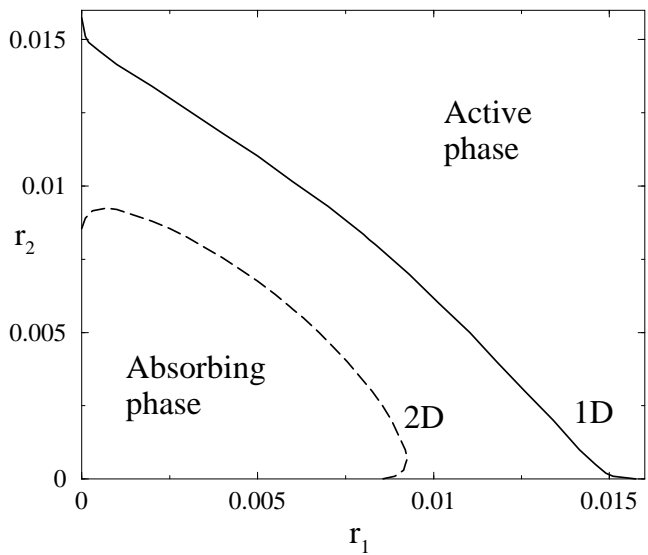}}
\caption{
\label{FIGGENERALIZED}
Phase diagram of the generalized model. The solid (dashed) line
indicates the phase transition line in one (two) dimensions.
}
\end{figure}

In order to support this point of view, we generalize 
the model in such a way that the continuous transition 
is shifted to the accessible region $r \geq 0$. 
This can be done by introducing two control parameters
$r_1,r_2$ and considering a site to be active if the 
product of adjacent bond variables lies in the interval $(-r_1,r_2)$.
Clearly, this model includes the original 
one as a special case. Moreover, the phase diagram is symmetric
under exchange $r_1 \leftrightarrow r_2$. In order to avoid
frozen pairs of sites, we will assume that both parameters are
positive. Obviously, for very small values of $r_1$ and $r_2$ the
spreading probability is so small that the model will be in the
absorbing phase. Increasing $r_1$ and $r_2$, we observe a continuous
transition from the absorbing to the active phase, as shown
in Fig.~\ref{FIGGENERALIZED}.

We verified that the critical behavior along the entire transition
line belongs to the universality class of directed percolation.
Note that the $Z_2$ symmetry along the diagonal $r_1=r_2$ does
not lead a different type of transition since it is not a symmetry
of the order parameter.

\section{Conclusions}

The common feature of the models introduced by Lipowski {\it et al.}
is the use of products of real-valued local variables in the definition
of the dynamic rules. The models display very interesting phenomena
which are usually not observed in models with linear local rules.
As we have shown in the present paper, the use of nonlinear rules is 
crucial since nonlinear functions of random numbers may not be 
uniformly distributed. A special situation emerges if the distribution 
exhibits a singularity. In this case a tiny change of the control 
parameter may lead to a dramatic variation of the order parameter. 
For the model investigated in~\cite{Lipowski}, such a singularity 
is responsible for the diverging slope in Fig.~\ref{FIGDENS}. 
Thus, in contrast to a  previous conjecture, the model does not 
combine features of first- and second-order phase transitions. 
In particular, the model is not critical and the correlation 
length remains finite as $r \rightarrow 0$.
We expect that similar phenomena may emerge whenever the dynamic rules
are defined in terms of nonlinear functions of real-valued random 
variables.


\end{document}